\documentclass[prl,twocolumn,amsmath,amssymb]{revtex4-1}
\usepackage{verbatim}
\usepackage{graphicx}% Include figure files
\usepackage{dcolumn}% Align table columns on decimal point
\usepackage{bm}% bold math
\usepackage[utf8]{inputenc}
\usepackage[T1]{fontenc}
\usepackage{mathptmx}
\usepackage{etoolbox}

\renewcommand{\r}{\vec{r}}

\newcommand{\ket}[1]{\left|#1\right>}
\newcommand{\bra}[1]{\left<#1\right|}
\newcommand{\bk}[2]{\left<#1|#2\right>}
\newcommand{\bak}[3]{\left<#1|#2|#3\right>}

\makeatletter
\def\@email#1#2{%
 \endgroup
 \patchcmd{\titleblock@produce}
  {\frontmatter@RRAPformat}
  {\frontmatter@RRAPformat{\produce@RRAP{*#1\href{mailto:#2}{#2}}}\frontmatter@RRAPformat}
  {}{}
}%
\makeatother
\begin{document}

\preprint{AIP/123-QED}

\title[CI-DFT+U]{DFT+U Type Strong Correlation Functional Derived from Multiconfigurational Wavefunction Theory} 
\author{Benjamin G. Janesko}
 \email{b.janesko@tcu.edu}

\affiliation{ Department of Chemistry \& Biochemistry, Texas Christian
University, Fort Worth, TX 76129, USA}

\date{\today}

\begin{abstract} 
We present a DFT+U-type functional for strong correlation, derived from
multiconfigurational wavefunction theory.  The reference system experiences
electron-electron interactions only in DFT+U-type atomic states, yielding a
block-localized configuration interaction Hamiltonian which depends on the
atomic state occupancies and the promotion energies of doubly excited
determinants.  Simple approximations for the promotion energies recover
the flat-plane condition and provide beyond-zero-sum accuracy for iron
spin-crossover complexes.
\end{abstract}

\maketitle 

\begin{figure}
\includegraphics[width=0.45\textwidth]{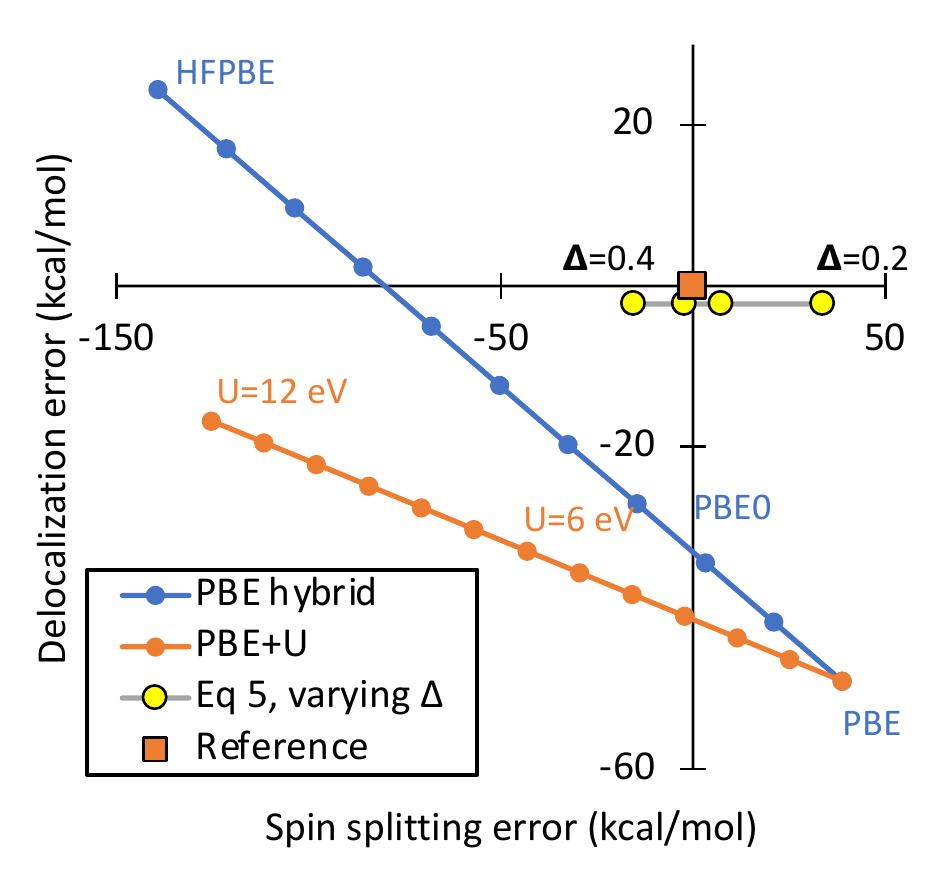} 
\caption{\label{fig:zero} (Abscissa) Error in singlet-quintet spin splitting of
[Fe(CO)$_6$]$^{2+}$, relative to CASPT2 reference.\cite{Finney2022} (Ordinate)
Energy difference per atom between localized (Fe$^{3+}\ldots$Fe$^{4+}$) and
delocalized (Fe$^{3.5+}\ldots$Fe$^{3.5+}$) ionizations of two asymptotically
separated high-spin Fe$^{3+}$ ions.} \end{figure}

Kohn-Sham density functional theory (DFT) introduces electron correlation into
a mean-field model of electronic structure. Standard DFT approximations do this
imperfectly, leading to zero-sum tradeoffs between over-delocalization of
charge and spin, and understimation of electron correlation.\cite{Janesko2021,Bryenton2022}
Figure \ref{fig:zero} shows an example of these tradeoffs, comparing the
delocalization error in fractionally charged Fe$^{(3+\delta)+}$, to the spin
splitting error in spin-crossover complex\cite{Finney2022} [Fe(CO)$_6$]$^{2+}$.
(Computational details are in the Supplemental Material). DFT+U and hybrid DFT 
approximations that accurately model the spin splitting introduce large delocalization
errors, and {\em{vice versa}}. Because of these tradeoffs, choosing (or tuning)
a DFT approximation to model one property often degrades accuracy for other
properties.\cite{Hughes2010} 
 
For decades, the DFT community has explored methods to go beyond these
tradeoffs. Many of these methods incorporate localized single-particle states to
capture aspects of delocalization \& correlation.\cite{Janesko2021} 
Examples include projection of the exchange hole onto a localized
model,\cite{Becke2005} comparisons of localized {\em{vs.}} delocalized exchange
holes,\cite{Perdew2008,Haasler2020} and projections of the reference system
wavefunction onto localized orbitalets,\cite{Su2018} atomic
states,\cite{Bajaj2017} or localized states at each point in
space.\cite{Verma2019}
Determining appropriate functional forms remains a challenge. For example, the
form of our nondynamical correlation model was based on educated guesses about
opposite-spin correlation.\cite{Ramos2020} Recent extended DFT+U methods employ
multi-parameter fits to recover the flat-plane
condition.\cite{Bajaj2017,Bajaj2019,Bajaj2022,Burgess2023}
This ambiguity can be contrasted with the generalized range-separated approach to
correcting DFT. In this approach,  part of the electron-electron interaction is
reintroduced into the reference system, the reference system wavefunction is
treated with a multideterminant approximation, and a modified
exchange-correlation (XC) density functional accounts for the remaining
electron-electron interactions.\cite{Savin1988,Autschbach2014}
The functional forms of generalized range-separated approaches are unambiguous,
free of double-counting, and can in principle be systematically improved to the
exact answer. 

This work present a DFT+U-type functional for strong correlation, derived
directly from a multiconfigurational wavefunction. The formalism is our further
generalization\cite{Janesko2022a} of the generalized range-separated approach.
The reference system's correlated wavefunction is modeled using ideas from
valence bond approaches.\cite{Beran2005} The XC functional is
modified\cite{Janesko2022b} following the Perdew-Zunger self-interaction
correction (PZSIC)\cite{Perdew1981} to eliminate double-counting. 
 
Our approach is based on the simplified rotationally averaged formulation of
DFT+U.\cite{Dudarev1998,Cococcioni2005,Kulik2015} In this approach, one
projects the reference system wavefunction onto $N_{proj}=N_{shell} L_{shell}$
orthonormal one-electron states $\{\tilde{\chi}_{Ip}\}$ localized at $N_{shell}$ atomic
shells $I$, and introduces an energetic penalty $U_{eff}$ for fractional
occupation. 
\begin{eqnarray}
\label{eq:DFTU}
E &=& E_{DFT} +  \sum_I\frac{U_{eff}}{2}\sum_{pq\sigma} \left(\tilde{n}^{I\sigma}_{pp}- \sum_q \tilde{n}^{I\sigma}_{pq}\tilde{n}^{I\sigma}_{qp}\right)  \\ 
&=& E_{DFT} +  \sum_I\frac{U_{eff}}{2}\sum_{p\sigma}\left(n_{Ip\sigma}-n_{Ip\sigma}^2\right)\nonumber
\end{eqnarray}
${\tilde{n}}^{I\sigma}_{pq}=\sum_i\bk{\psi_{i\sigma}}{\tilde{\chi}_{Ip}}\bk{\tilde{\chi}_{Iq}}{\psi_{i\sigma}}$
is the $\sigma$-spin projected density matrix obtained by
projecting the reference system single-determinant wavefunction
$\ket{\Phi_0}$, built from occupied spinorbitals $\{\psi_{i\sigma}\}$, onto 
$\{\tilde{\chi}_{Ip}\}$.  $n_{Ip\sigma}$ is the $p$th eigenvalue of the
$L_{shell} x L_{shell}$ block ${\bf{\tilde{n}}}^{I\sigma}$, corresponding to
eigenfunction $\chi_{Ip\sigma}$. 
(We assume that ${\bf{{\tilde{n}}}}$ is block-diagonal in spin $\sigma$ and
shells $I$.)
Extended DFT+U approaches add terms such as $n_{pq}^{I\uparrow} n_{qp}^{I\downarrow}$
to model strong correlation.\cite{Bajaj2017,Bajaj2019,Bajaj2022,Burgess2023}

The present work extends our rederivation of DFT+U,\cite{Janesko2022b} which
builds on the connections between DFT+U and hybrid DFT.\cite{Andriotis2010,Ivady2014,Agapito2015} Consider an $N$-electron system in
external potential $\hat{V}_{ext}$ and a reference system of $N$ noninteracting
electrons whose single-determinant ground state wavefunction $\ket{\Phi_0}$
yields density $\rho(\r)=\sum_{i\sigma}|\psi_{i\sigma}(\r)|^2$. 
%
%The reference system Hamiltonian is $\hat{T}+\hat{V}_{ext}$. 
%
The ground-state energy is 
%
%\begin{eqnarray} \label{eq:dft}
$E_{DFT} = \bak{\Phi_0}{\hat{T}+\hat{V}_{ext}}{\Phi_0} +
%\sum_{i\sigma}\bak{\psi_{i\sigma}}{\hat{T}+\hat{V}_{ext}}{\psi_{i\sigma}} +
J[\rho] + E_{XC}[\rho_\uparrow,\rho_\downarrow]$. 
%\end{eqnarray}
%
%Here $\hat{T}(i)$ is the electron kinetic energy operator acting on electron
%$i$. 
$\hat{T}$ and $\hat{V}_{ext}$ are kinetic and external potential
operators. $J[\rho]$ is the mean-field Hartree energy corresponding to density
$\rho$.  The ``exact'' exchange piece of $E_{XC}$,
$E_X^{ex}[\rho]=\bak{\Phi_0}{\hat{V}_{ee}}{\Phi_0}-J[\rho]$, is defined in
terms of the electron-electron interaction operator
$\hat{V}_{ee} = \sum_{i>j}|\r_i-\r_j|^{-1}$
%
%whih acts all pairs of electrons $i,j$. 
%
The Hohenberg-Kohn theorems ensure that minimizing $E_{DFT}$ using the exact
$E_{XC}[\rho_\uparrow,\rho_\downarrow]$ recovers the exact ground-state density
and energy of the real system. 
 
To rederive DFT+U, we project $\hat{V}_{ee}$ onto the
states $\{\chi_{Ip\sigma}\}$ introduced above: 
\begin{eqnarray}
\label{eq:VP}
\hat{V}^P_{ee} &=& \sum_{i>j} \sum_{Ip\sigma} \ket{\chi_{p\sigma}(i)
\chi_{Ip\sigma}(j)}J_I\bra{\chi_{Ip\sigma}(i)\chi_{Ip\sigma}(j)}
\end{eqnarray}
$J_{I}=\left<\chi_{Ip\sigma}(1)\chi_{Ip\sigma}(2)||\r_1-\r_2|^{-1}|\chi_{Ip\sigma}(1)\chi_{Ip\sigma}(2)\right>$
is the self-Hartree interaction of an electron in state $\chi_{Ip\sigma}$.
%$\ket{\chi_{Ip\sigma}(i)}$ denotes projection onto electron $i$. 
%assumed to be the same for all states in shell $I$. 
%
The reference system Hamiltonian becomes $\hat{T}+\hat{V}_{ext}+\hat{V}^P_{ee}$. 
Because $\{\chi_{Ip\sigma}\}$ are eigenfunctions of the projected
one-particle density matrix corresponding to $\ket{\Phi_0}$, the projection 
$\sum_i\bk{\Phi_0}{\chi_{Ip\sigma}(i)}\bk{\chi_{Iq\sigma}(i)}{\Phi_0}=n_{Ip\sigma}\delta_{pq}$. 
This ensures that the expectation value of the projected electron-electron interaction 
$\bak{\Phi_0}{\hat{V}_{ee}^P}{\Phi_0}$ becomes $\sum_I (J_{I}/2)\sum_{p\sigma}
(n_{Ip\sigma}^2-n_{Ip\sigma}^2)$,  which equals zero. Thus, $\hat{V}^P_{ee}$ 
introduces only electron self-interaction into
the reference system. The reference system's exact ground-state wavefunction is
a single Slater determinant, and the Hohenberg-Kohn theorems ensure the
existence of an exact projected Hartree-exchange-correlation density
functional. We define the projected Hartree functional as
$J^P[\rho]=J[\rho]-\sum_{Ip\sigma}J_{I}n_{Ip\sigma}^2$. Inspired by the
PZSIC,\cite{Perdew1981} we model the projected XC functional by passing
projected single-particle densities $n_{Ip\sigma}|\chi_{Ip\sigma}(\r)|^2$ to 
the XC functional, giving 
$E_{XC}^P[\rho_\uparrow,\rho_\downarrow]=E_{XC}[\rho_\uparrow,\rho_\downarrow]-\sum_{Ip\sigma}
E_{XC}[n_{Ip\sigma}|\chi_{Ip\sigma}|^2,0]$. The ground-state energy becomes 
\begin{eqnarray}
\label{eq:USIC}
E &=& E_{DFT} - \sum_I \frac{J_I}{2}\sum_{p\sigma} 
\left(n_{Ip\sigma}^2+2J_{I}^{-1}E_{XC}[n_{Ip\sigma}|\chi_{Ip\sigma}|^2,0]
\right)
\end{eqnarray}
(The projected Hartree energy is included in the Hartree energy in $E_{DFT}$.)
Approximating the double-counting correction as
$E_{XC}[n_{Ip\sigma}|\chi_{Ip\sigma}|^2,0]\simeq -c_1 n_{Ip\sigma}-c_2
n^2_{Ip\sigma}$, constrained to obey $c_1=(J_I-c_2)$, makes eq \ref{eq:USIC}
recover eq \ref{eq:DFTU}.  The resulting $U_{eff}=(J_I-c_2)$ has a transparent
dependence on the choice of projection states\cite{Andriotis2010} and
approximate XC functional. The exact XC functional, or any functional which is
exact for one-electron systems, correctly gives $U_{eff}=0$.

We next add to $\hat{V}_{ee}^P$ the opposite-spin interactions $\sum_{i>j}
\sum_{Ip} \sum_{\sigma'\neq \sigma} \ket{\chi_{Ip\sigma}(i)
\chi_{Ip\sigma'}(j)}J_I\bra{\chi_{Ip\sigma}(i)\chi_{Ip\sigma'}(j)}$. 
(We assume negligible spin contamination of $\tilde{{\bf{n}}}^{I\sigma}$ so
that each eigenfunction $\chi_{Ip\uparrow}$ corresponds to one $\chi_{Ip\downarrow}$.) 
The projected Hartree and XC functionals become 
$J^P[\rho]=J[\rho]-\sum_{Ip\sigma\sigma'}J_{I}n_{Ip\sigma}n_{Ip\sigma'}$ and
$E_{XC}^P[\rho_\uparrow,\rho_\downarrow]=E_{XC}[\rho_\uparrow,\rho_\downarrow]-\sum_{Ip}
E_{XC}[n_{Ip\uparrow}|\chi_{Ip\uparrow}|^2,n_{Ip\downarrow}|\chi_{Ip\downarrow}|^2]$. 
$\bak{\Phi_0}{\hat{V}^P_{ee}}{\Phi_0}$ is no longer zero and the reference
system's exact ground-state wavefunction is no longer a single Slater
determinant. 

We approximate this correlated wavefunction using ideas from valence bond
theory.\cite{Beran2005}  We perform a unitary transform of the occupied
orbitals $\{\psi_{i\sigma}\}$ in $\ket{\Phi_0}$, so that each $\chi_{Ip\sigma}$
has a nonzero projection onto only one transformed occupied orbital denoted
$\phi^{(o)}_{Ip\sigma}$.  The unitary transform does not change the projection
of $\ket{\Phi_0}$ onto state $\chi_{Ip\sigma}$, ensuring
$|\bk{\chi_{Ip\sigma}}{\phi^{(o)}_{I'p'\sigma'}}|^2=n_{Ip\sigma}\delta_{II'}\delta_{pp'}\delta_{\sigma\sigma'}$.
We perform a corresponding transform of the unoccupied (virtual) Kohn-Sham
orbitals $\{\psi_{a\sigma}\}$, such that each $\chi_{Ip\sigma}$ has a nonzero
projection onto only one transformed virtual orbital $\phi^{(v)}_{Ip\sigma}$.
Orthogonality of occupied-virtual spaces and completeness of the
occupied+virtual space ensures
$|\bk{\chi_{Ip\sigma}}{\phi^{(v)}_{Ip\sigma}}|^2=1-n_{Ip\sigma}$. 
(A key to our approach is that these transforms can be either
approximated, or evaluated explicitly.) 
We approximate the reference system's wavefunction as a superposition of
$\ket{\Phi_0}$ and the $N_{proj}$ ``doubly excited'' Slater determinants
$\ket{\Phi_{Ipo}^{Ipv}}$ in which $\phi^{(o)}_{Ip\uparrow}$ is replaced by
$\phi^{(v)}_{Ip\uparrow}$ and $\phi^{(o)}_{Ip\downarrow}$ is replaced by
$\phi^{(v)}_{Ip\downarrow}$. 
$\hat{V}^P_{ee}$ does not include electron-electron interactions between the
$N_{proj}$ different states $p$. (These interactions are treated by the
projected Hartree-exchange-correlation functional.) Accordingly, the reference
system's configuration interaction (CI) Hamiltonian is block-diagonalized into
$N_{proj}$ $2x2$ matrices 
\begin{eqnarray}
\label{eq:CI}
\left(\begin{array}{cc} 0 &  \bak{\Phi_0}{\hat{V}_{ee}^P}{\Phi_{Ipo}^{Ipv}} \\
%J_{I}\sqrt{n_{Ip\uparrow}(1-n_{Ip\uparrow})n_{Ip\downarrow}(1-n_{Ip\downarrow})} 
\bak{\Phi_{Ipo}^{Ipv}}{\hat{V}_{ee}^P}{\Phi_0} & J_I\Delta_{Ip} \end{array}\right) 
\end{eqnarray}
Here
$\bak{\Phi_0}{\hat{V}_{ee}^P}{\Phi_{Ipo}^{Ipv}}$
$=\bak{\phi^{(o)}_{Ip\uparrow}\phi^{(o)}_{Ip\downarrow}}{\hat{V}^P_{ee}}{{\phi^{(v)}_{Ip\uparrow}\phi^{(v)}_{Ip\downarrow}}}$
equal 
$J_{I}\sqrt{n_{Ip\uparrow}(1-n_{Ip\uparrow})n_{Ip\sigma'}(1-n_{Ip\downarrow})}$. 
The diagonal elements introduce the unitless promotion energy
$\Delta_{Ip}$=$J_{I}^{-1}\left(\bak{\Phi_{Ipo}^{Ipv}}{\hat{H}_0}{\Phi_{Ipo}^{Ipv}}
- \bak{\Phi_0}{\hat{H}_0}{\Phi_0}\right)$, where $\hat{H}_0$ is the Hamiltonian
  of the projected-interacting reference system.  $\Delta_{Ip}\geq 0$,
otherwise $\ket{\Phi_0}$ would not be the reference system's lowest-energy
single Slater determinant wavefunction. 
The reference system's correlation energy is the sum of the lower eigenvalues
of these $2x2$ Hamiltonian blocks. The final energy is 
\begin{eqnarray}
\label{eq:UGVB}
E &=& E_{DFT} + \sum_I \frac{J_I}{2}\sum_p \Big(-\sum_\sigma n_{Ip\sigma}^2 \\
&+& \Delta_{Ip}-\left(\Delta_{Ip}^2+4n_{Ip\uparrow}n_{Ip\downarrow}(1-n_{Ip\uparrow})(1-n_{Ip\downarrow})\right)^{1/2}
\nonumber \\ 
&-& 2J_I^{-1}E_{XC}[n_{Ip\uparrow}|\chi_{Ip\uparrow}|^2,n_{Ip\downarrow}|\chi_{Ip\downarrow}|^2]\Big) \nonumber 
\end{eqnarray}
%
%The double excitation energy (the difference between the CI eigenvalues), is
%$J\left(\Delta^2+4n_{Ip\uparrow}n_{Ip\downarrow}(1-n_{Ip\uparrow})(1-n_{Ip\downarrow})\right)^{1/2}$. 
%
Eq \ref{eq:UGVB} does not double-count exchange or correlation, just as
the PZSIC does not double-count exchange or correlation. 
Eq \ref{eq:UGVB} is reminiscent of the forms of other extended DFT+U
approximations.\cite{Bajaj2017,Bajaj2019,Bajaj2022,Burgess2023}

We may evaluate eq \ref{eq:UGVB} exactly for asymptotically stretched
singlet hydrogen molecule H$_2$ in an orthonormal minimal basis set of
functions $\chi_A,\chi_B$ localized to atoms A and B.  We construct orthonormal
spinorbitals $\psi_{1\sigma}=c_A\chi_{A\sigma}+(1-c_A^2)^{1/2}\chi_{B\sigma}$,
$\psi_{2\sigma}=(1-c_A^2)^{1/2}\chi_{A\sigma}-c_A\chi_{B\sigma}$ . 
The spin-symmetry-unrestricted mean-field state
$\ket{\Phi_0}=\ket{\psi_{1\uparrow}\psi_{2\downarrow}}$ places
$n_{A\uparrow}=|c_A|^2$ $\uparrow$-spin electrons and
$n_{A\downarrow}=1-n_{A\uparrow}$ $\downarrow$-spin electrons on atom A. 
For any $n_{A\uparrow}$, configurations
$\ket{\psi_{1\uparrow}\psi_{2\downarrow}}$ and
$\ket{\psi_{2\uparrow}\psi_{1\downarrow}}$ are degenerate. 
For the exact ground state, all configurations with 
$0\leq n_{A\uparrow}\leq 1$ are degenerate, the so-called fractional spin line
(FSL).\cite{Yang2000} Unrestricted Hartree-Fock calculations give the exact
energy at $n_{A\uparrow}=0$ or 1, and insufficiently negative energies at
intermediate $n_{A\uparrow}$ (fractional spin error).
To evaluate eq \ref{eq:UGVB}, we project the electron-electron interaction operator onto
$N_{shell}=2$ shells each with $L_{shell}=1$, {\em{i.e.}}, onto $\chi_A$ and $\chi_B$.
Because $L_{shell}=1$, the projected density matrix eigenfunction
$\chi_{I=1,p=1,\sigma}$ introduced above is $\chi_{A\sigma}$. 
The projected XC energy $E_{XC}^P =
E_{XC}[\rho]-\sum_{S=A,B}E_{XC}[n_{S\uparrow}|\chi_S|^2,n_{S\downarrow}|\chi_S|^2]$ is zero. 
Because there is only one occupied and one virtual orbital of each spin, the
projected states in eq \ref{eq:CI} become 
$\phi^{(o)}_{I=1,p=1,\sigma}=\psi_{1\sigma}$ and
$\phi^{(v)}_{I=1,p=1,\sigma}=\psi_{2\sigma}$, such that
$\ket{\Phi_{Ipo}^{Ipv}} = \ket{\psi_{2\uparrow}\psi_{1\downarrow}}$.  Because
$\ket{\psi_{2\uparrow}\psi_{1\downarrow}}$ and
$\ket{\psi_{1\uparrow}\psi_{2\downarrow}}$ are degenerate, $\ket{\Phi_0}$ and
$\ket{\Phi_{Ipo}^{Ipv}}$ are degenerate and $\Delta$=0. Eq \ref{eq:CI} with
$\Delta=0$ becomes the full configuration interaction (FCI) Hamiltonian in this
minimal basis, and eq \ref{eq:UGVB} is the exact FCI ground-state energy. 

\begin{figure}
\includegraphics[width=0.45\textwidth]{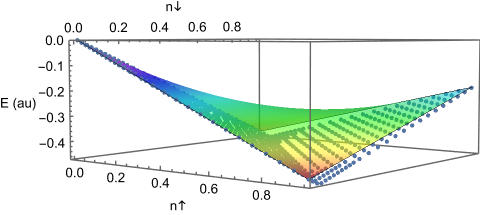}
\caption{\label{fig:fp} Eq \ref{eq:UGVB} (points) and eq \ref{eq:USIC} (surface)
for minimal-basis hydrogen atom with occupancy $n_\uparrow$ and
$n_\downarrow$.}
\end{figure} 

Eq \ref{eq:UGVB} can be evaluated for the entire ``flat-plane condition''. We
follow ref \citenum{MoriSanchez2014}, treating an integer number of electrons
distributed across asymptotically separated hydrogen nuclei such that each atom
has average populations $0\leq n_\uparrow,n_\downarrow\leq 1$.  The exact
energy of the real system is (to within infintesimal error) a sum of subsystem
energies $E[n_\uparrow,n_\downarrow]$, where $E[n_\uparrow,n_\downarrow]$ is
composed of two flat planes intersecting at the
FSL.\cite{MoriSanchez2014,Kong2022} 
Figure \ref{fig:fp} illustrates eq \ref{eq:USIC} and \ref{eq:UGVB} for this
system. Calculations use the STO-6G basis set, the Perdew-Burke-Ernzerhof (PBE)
XC functional,\cite{Perdew1996} and choose the projection state as the STO-6G
basis function for hydrogen atom. 
PySCF code\cite{Sun2020} for the calculations in Figure \ref{fig:fp} is
provided as Supplementary Information.
We consider eq \ref{eq:UGVB} with
$\Delta=|n_\uparrow+n_\downarrow-1|\equiv \Delta_{FP}$. 
These calculations recover the exact flat-plane energy,\cite{MoriSanchez2014}
including constant energy along the FSL, zero correlation energy for the
minimal-basis closed-shell anion $n_\uparrow=n_\downarrow=1$, and a
discontinuity when crossing the FSL. 
This approach is nearly exact for other atoms and larger basis sets
(Supplementary Information). 
The form of $\Delta_{FP}$ may be rationalized as follows. When
$n=n_\uparrow+n_\downarrow$ is 1, $\Delta=0$ as derived above. As $n$
approaches 0 or 2, the Hartree-Fock energy becomes an increasingly good
approximation to FCI. the FCI wavefunction is increasingly dominated by a
single determinant, thus $\Delta$ must increase. 
The correction to $E_{DFT}$, plotted as a function of $n_\uparrow,n_\downarrow$, is
qualitatively similar to other extended DFT+U methods (Supplementary
Information). 
%
%Figure \ref{fig:jmDFT} shows the correction to $E_{PBE}$ provided by eq
%\ref{eq:UGVB}. The sum of exact exchange
%$-(J/2)(n_{\uparrow}^2+n_{\downarrow}^2)$, opposite-spin correlation from eq
%\ref{eq:CI}, and the DFT correction
%$E_{PBE}[n_\uparrow|\chi|^2,n_\downarrow|\chi|^2]$, effectively recovers the
%judiciously modified DFT correction in Figure 2 of ref \cite{Bajaj2017}. 
%

\begin{table}
\begin{tabular*}{0.44\textwidth}{@{\extracolsep{\fill}}l rrrrr}
\hline
\hline
Ligand L & NH$_3$& NCH & CO & CNH & MAE \\ 
\hline
Ref\cite{Finney2022} & -41.5 & -16.7 & 37.6 & 54.7 & 0.0\\ 
PBE & -1.6 & 18.4 & 76.4 & 88.8 & 37.0 \\ 
PBE0 & -19.5 & -8.7 & 31.9 & 45.8 & 11.1 \\ 
PBE+U & -37.4 & -32.2 & -5.7 & 6.9 & 27.7 \\ 
Eq \ref{eq:USIC} & -82.7 & -137.8 & -177.6 & -154.7 & 146.7 \\ 
Eq \ref{eq:UGVB}, $\Delta_{FP}$ & -142.1& -167.6 & -91.7 & -53.5 & 122.3 \\ 
Eq \ref{eq:UGVB}, $\Delta=0.35$  & -35.5 & -47.6 & 22.1 & 53.2 & 13.5 \\ 
\hline
\end{tabular*}
\caption{\label{tab:Fe} Predicted quintet-singlet splittings of
[FeL$_6$]$^{2+}$ spin crossover complexes (kcal/mol), positive values denote
stable singlet, and mean absolute error (MAE).\cite{Finney2022} }
\end{table}

\begin{figure}
\includegraphics[width=0.44\textwidth]{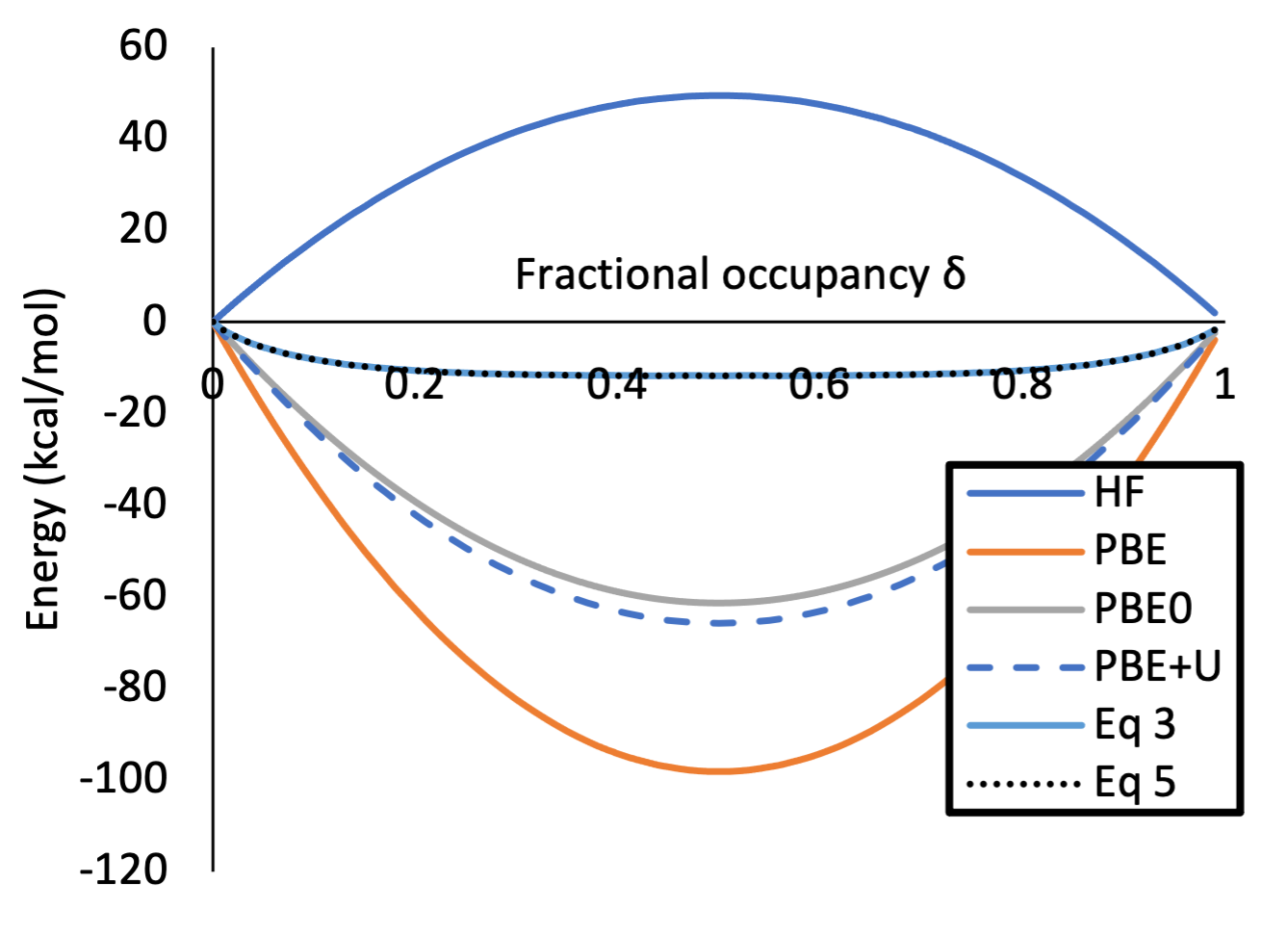}
\caption{\label{fig:Fe} Ionization energy of high-spin Fe$^{3+}\ldots$Fe$^{3+}$
to Fe$^{(3+\delta)+}\ldots$Fe$^{(4-\delta)+}$, relative to energy at
$\delta=0$. Eq \ref{eq:USIC} and eq \ref{eq:UGVB} give nearly identical
results.}
\end{figure}

Eq \ref{eq:UGVB} can be applied to more realistic systems.  Table \ref{tab:Fe}
shows the predicted spin splitting of four iron(II) spin-crossover (SCO)
complexes.  Figure \ref{fig:Fe} illustrates delocalization error, computed
as the energy to ionize a pair of distant high-spin Fe$^{3+}$ ions to yield
(Fe$^{(3+\delta)+}\ldots$Fe$^{(4-\delta)+}$), relative to the energy at
$\delta=0$. 
Figure \ref{fig:zero} plots the predicted spin splitting of [Fe(CO)$_6$]$^{2+}$
versus the delocalization error at $\delta=1/2$.
Eq \ref{eq:USIC} and eq \ref{eq:UGVB} both include full exact exchange in the
projected states, and both give low delocalization error in Figure \ref{fig:Fe}.
However, eq \ref{eq:USIC} gives underestimated spin splittings in
Table \ref{tab:Fe}, and eq \ref{eq:UGVB} with $\Delta_{FP}$ 
The correlation terms in eq \ref{eq:UGVB} correct these underestimations.
While $\Delta_{FP}$ is exact for the flat-plane condition, it is insufficiently
robust for practical calculations (Supplementary Information) and
underestimates the spin splittings in Table \ref{tab:Fe}. 
Choosing constant $\Delta=0.35$ accurately models spin splittings.
More importantly, Figure \ref{fig:zero} shows that different choices of
$\Delta$ all give negligible delocalization error for Fe$^{3+}$ ionization.  Eq
\ref{eq:UGVB} provides {\bf{beyond-zero-sum}} performance at mean-field cost. 

The results presented here provide a new perspective on beyond-zero-sum DFT. By
projecting the electron-electron interaction onto specific atomic states, we
can introduce a controlled degree of electron correlation into the reference
system. Exact or nearly exact treatments of this strong correlation may be
rigorously embedded into DFT.  This general derivation may be extended to other
single-particle states and
more sophisticated approximations for the promotion energy $\Delta$, providing
new insight into DFT-based treatments of strong
correlation.\cite{Bajaj2017,Bajaj2019,Bajaj2022,Burgess2023}

\bibliographystyle{aip}
%\bibliography{../../../review/tosubmit/tosubmit}

%%%%%%%%%%%%%%%%%%%%%%%%%%%
\clearpage\newpage
\widetext
\begin{center}
\textbf{\large Supplemental Materials: 
DFT+U Type Strong Correlation Functional Derived from Multiconfigurational Wavefunction Theory} 
\end{center}
%%%%%%%%%%% Merge with supplemental materials %%%%%%%%%%
%%%%%%%%%%% Prefix a "S" to all equations, figures, tables and reset the counter %%%%%%%%%%
\setcounter{equation}{0}
\setcounter{figure}{0}
\setcounter{table}{0}
\setcounter{page}{1}
\makeatletter
\renewcommand{\theequation}{S\arabic{equation}}
\renewcommand{\thefigure}{S\arabic{figure}}
\renewcommand{\bibnumfmt}[1]{[S#1]}
\renewcommand{\citenumfont}[1]{S#1}

\section{Computational Details }

Eq 3 and eq 5 in the manuscript are evaluated non-self-consistently using
code provided below.  
To simplify the implementation, all calculations take the projection states
from a minimal atomic basis set. 
To illustrate, consider a def2TZVP calculation on Fe(CO)$_6$. First, we project
the one-particle density matrix onto the minimal STO-6G basis set for the entire
complex. Next, we extract the 5x5 block of the density matrix corresponding to
the five 3d atomic orbitals. We construct a new density matrix for isolated
STO-6G iron atom, which equals the 5x5 projected density matrix for the 3d
orbitals, and equals 0 elsewhere. We pass this projected density matrix to a
standard DFT integration routine for the minimal-basis iron atom. 
Calculations in Figure 2 treat a single hydrogen atom with a single STO-6G
basis function $\chi$, and one-particle density matrices
$\gamma_\uparrow=\ket{\chi}n_\uparrow\bra{\chi}$ and
$\gamma_\downarrow=\ket{\chi}n_\downarrow\bra{\chi}$. 
Calculations on the spin-crossover complexes use the TPSSh complex geometries
reported in ref 3, the reference CASPT2 spin splittings
reported at those geometries, and PBE0/def2tzvp orbitals and one-particle
density matrices computed with the Gaussian 16 package. Calculations are
performed spin-symmetry-restricted for the low-spin complexes and
spin-unrestricted for the high-spin complexes. Spin contamination is low, with
a maximum $\left<S^2\right>=6.0237$. 
Calculations on the distant Fe dimers model the one-particle density matrix of
Fe$^{(3+\delta)+}$ as a weighted sum of the spin-unrestricted high-spin
PBE0/def2tzvp density matrices of Fe$^{3+}$  and Fe$^{4+}$. 

\section{Numerical Issues of $\Delta_{FP}$}

\begin{figure}[ht]
\includegraphics[width=0.45\textwidth]{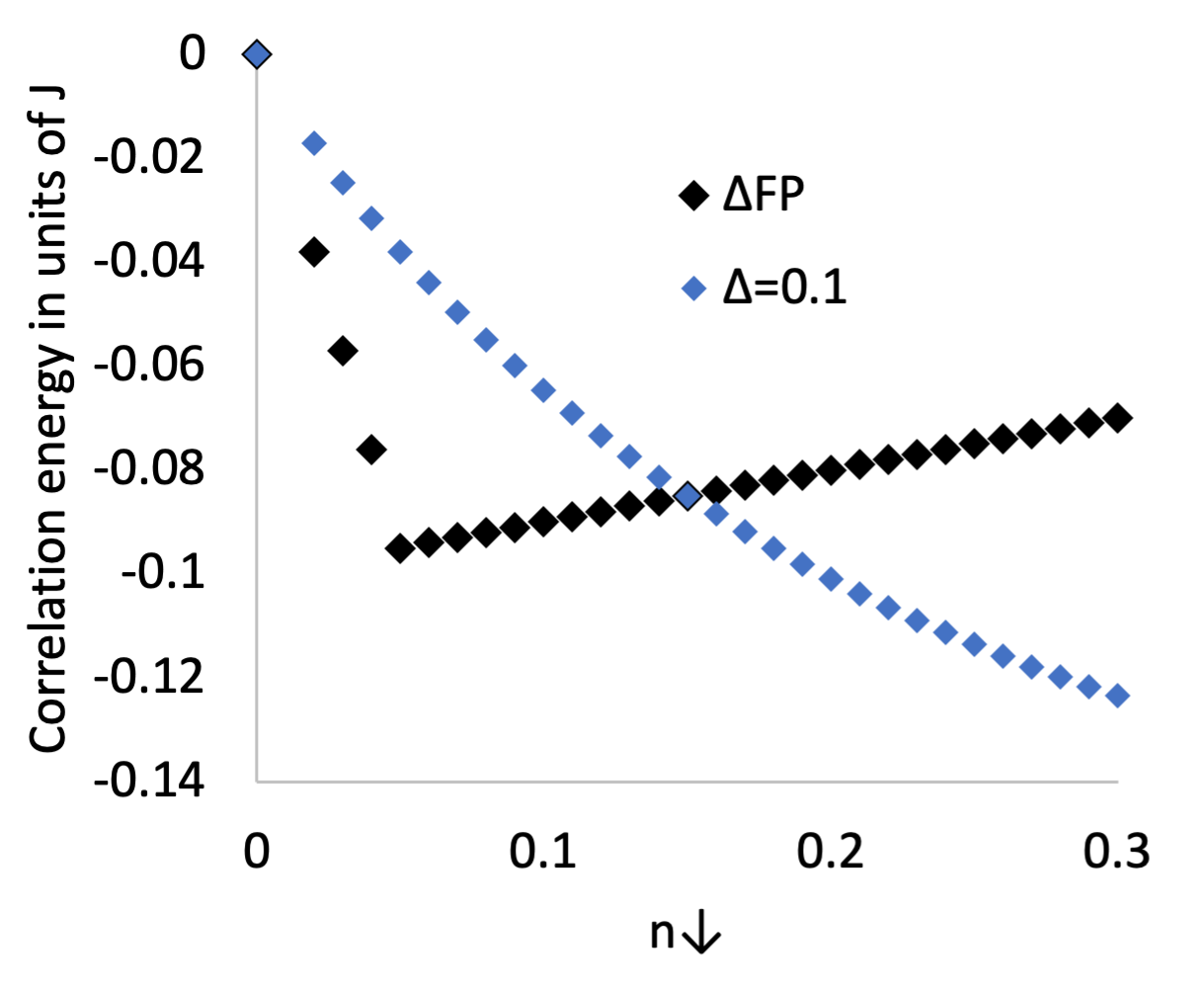}
\caption{\label{fig:DeltaFP} Correlation energy of eq 3 evaluated
at $n_\uparrow=0.95$, plotted vs. $n_\downarrow$.}
\end{figure}

While $\Delta_{FP}$ can exactly recover the flat-plane condition, it is
insufficiently robust for practical calculations. In spin-polarized systems,
the majority-spin projection $n_{Ip\uparrow}$ is often near but not quite 1.
As the minority-spin projection $n_{Ip\downarrow}$ increases from zero,
$\Delta_{FP}=|n_{Ip\uparrow}+n_{Ip\downarrow}-1|$ passes through a minimum,
producing a discontinuity in the opposite-spin correlation energy.  Figure
\ref{fig:DeltaFP} illustrates an example for $n_{Ip\uparrow}=0.95$. 

\section{The Flat-Plane Condition For Other Atoms And Basis Sets}

\begin{figure}[ht]
\includegraphics[width=0.45\textwidth]{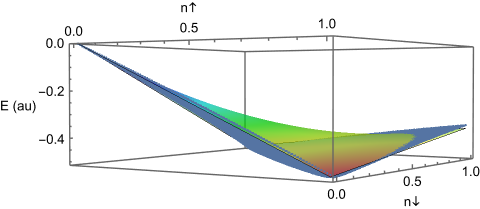}
\includegraphics[width=0.45\textwidth]{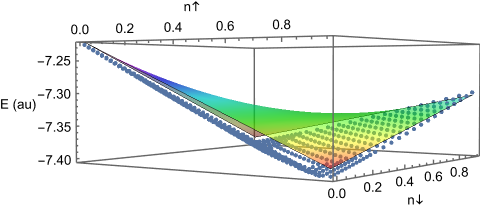}
\caption{\label{fig:fpqz} Eq 5 (points) and eq 3 (surface)
for def2QZVP-basis hydrogen atom (left) and minimal basis lithium atom (right).}
\end{figure}

Figure 2 shows the flat-plane condition for hydrogen atom, computed with
the def2QZVP basis set, keeping the STO-6G basis function as the projection
state, and choosing $\Delta=|n_\uparrow+n_\downarrow-n_{max}|$. 
$n_{max}$=0.96597 is the maximum value of $n_\sigma$ for this system,
{\em{i.e.}}, the projection of the spin-polarized UHF/def2QZVP ground
state onto the STO-6G projection state. (For the calculations in Figure
2, the minimal basis ground state perfectly overlaps with the
minimal basis projection state and $n_{max}=1$.) 
This mismatch means that the projected XC energy in eq 5 is no
longer zero, as it was in Figure 2. The remaining XC contributions
produce a slight nonplanarity in the predicted total energies.

\section{Other Supplementary Figures} 

\begin{figure}[ht]
\includegraphics[width=0.45\textwidth]{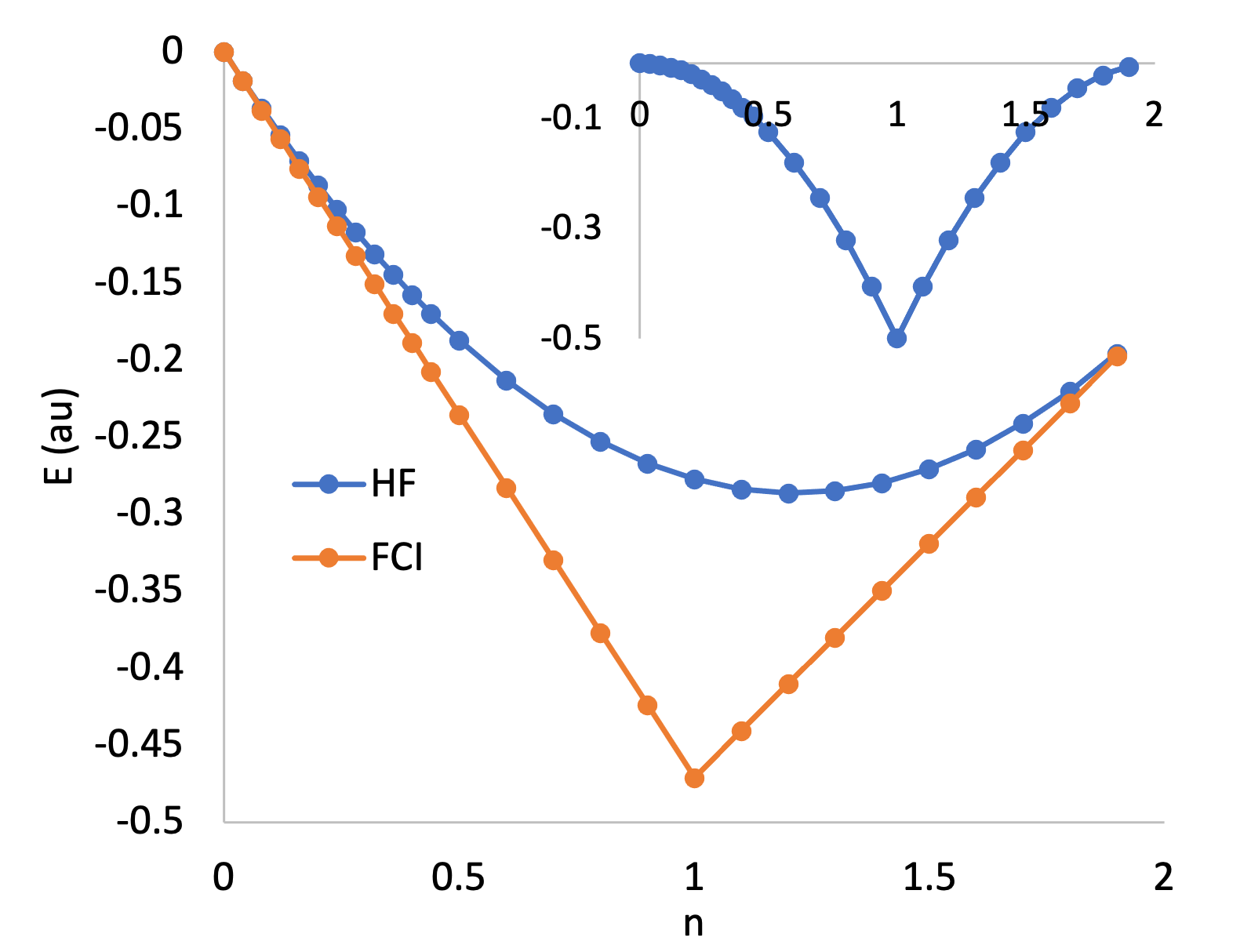}
\caption{\label{fig:diag} Hartree-Fock and exact energies of minimal basis H
atom with $n_\uparrow=n_\downarrow$ electrons, plotted vs
$n=n_\uparrow+n_\downarrow$. Inset shows the correlation energy in units of $J$
(points), and the fit -$J/2\min{n^2,(2-n)^2}$ (lines).}
\end{figure}

\begin{figure}[ht]
\includegraphics[width=0.45\textwidth]{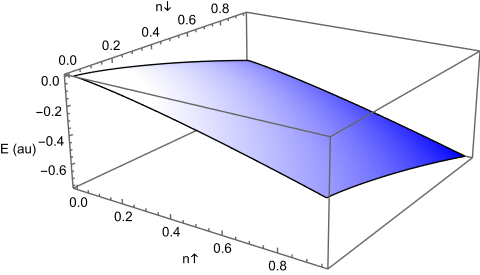}
\includegraphics[width=0.45\textwidth]{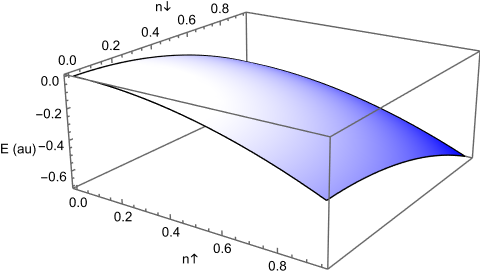}
\includegraphics[width=0.45\textwidth]{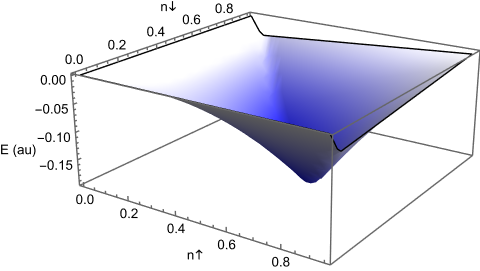}
\includegraphics[width=0.45\textwidth]{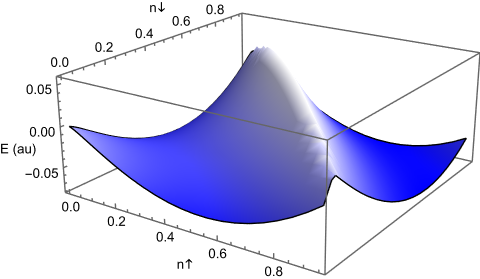} \caption{\label{fig:jmDFT}
Corrections to the flat-plane condition in Figure 2. (Top left)
Projected XC correction $E_{XC}^{PBE}[n_\uparrow|\chi_A|^2,n_\downarrow|\chi_A|^2]$.
(Top right) Exact exchange $-J/2(n_\uparrow^2+n_\downarrow^2)$. (Bottom left)
Opposite-spin correlation. (Bottom right) Summed
corrections.}
\end{figure} 

\clearpage\newpage 
\section{PySCF code for Figure 2} 

{\tiny
\begin{verbatim}
from pyscf import scf,gto,dft
import numpy ,scipy

xc='pbe,pbe'
a2e=27.211

# Computing the flat-plane condition in H atom 
nproj=1 
tag='s'
#tag='2s'
m=gto.Mole(atom='H',basis='def2qzvp',charge=0 ,spin=1) 
mmin=gto.Mole(atom='H',basis='sto-6g',charge=0 ,spin=1)
matmin=gto.Mole(atom='H',basis='sto-6g',charge=0 ,spin=1)
m.build()
mmin.build()
matmin.build()

# Function to evaluate the DFT correction EXC[na*rhop,nb*rhop] in the projection states 
matd=dft.UKS(matmin,xc=xc)
S = matd.get_ovlp()
Nat=S.shape[0]
def EXCp(na,psia,nb,psib):
  P = numpy.zeros((2,Nat,Nat))
  np1 = Nat-nproj
  P[0,np1:Nat,np1:Nat]=na*numpy.einsum('i,j->ij',psia,psia)
  P[1,np1:Nat,np1:Nat]=nb*numpy.einsum('i,j->ij',psib,psib)
  n,exc,vxc=matd._numint.nr_uks(mol=matmin,grids=matd.grids,xc_code=matd.xc,dms=P)
  return(exc)


Uval = 6/27.211/2. # Fe uses U = 6 eV

# Compute J from the Hartree energy of the isolated atom with one occupied orbital 
P0temp = numpy.zeros((Nat,Nat))
P0temp[Nat-1,Nat-1] = 1
J = 0.5*numpy.einsum('ij,ij->',P0temp,matd.get_j(dm=P0temp))
print('The U and J values: ',Uval,J)

# Build cross-overlap projector 
S = m.intor_symmetric('int1e_ovlp')
Sm = numpy.linalg.inv(S)
SM = mmin.intor_symmetric('int1e_ovlp')
SX = gto.intor_cross('int1e_ovlp',m,mmin)
      
# Find the minimal basis AOs in the molecule, these will be the DFT+U projection functions 
daos=[]
labs = mmin.ao_labels()
for iao in range(mmin.nao):
  icen = int(labs[iao].split()[0])
  iat = mmin.atom_charge(icen)
  if(tag in labs[iao] ):
    daos.append(iao)

SXP = numpy.zeros((m.nao,nproj))
SSP = numpy.zeros((nproj,nproj))
ind = 0 
for i in daos:
  SXP[:,ind] = SX[:,i]
  ind2 = 0 
  for j in daos:
    SSP[ind,ind2] = SM[i,j]
    ind2 = ind2+1 
  ind = ind+1 

# Lowdin orthogonalization of the projection states 
(SSvals,SSvecs) = numpy.linalg.eigh(SSP)
smhalf=numpy.identity(nproj)*0.0 
for i in range(nproj): 
  if(SSvals[i]>0.0000001):
     smhalf[i,i] = SSvals[i]**(-0.5)
SSvecs = numpy.dot(SSvecs,numpy.dot(smhalf,numpy.transpose(SSvecs)))
SSVoverlap = numpy.dot(SSvecs,numpy.dot(SSP,numpy.transpose(SSvecs)))
SXPo =numpy.dot(SXP,numpy.transpose(SSvecs))
Stest = numpy.einsum('im,ij,jn->mn',SXPo,Sm,SXPo)

# Build nprojxnproj projected 1PDM from full-basis molecule 1PDM
def ProjDM(P):
  Pproj = numpy.einsum('im,sij,jn->smn',SXPo,P,SXPo)
  return(Pproj)

# Convert nprojxnproj projected potential into full-basis potential 
def UnprojPot(Vproj):
  V = numpy.einsum('ik,km,smn,ln,lj->sij',Sm,SXPo,Vproj,SXPo,Sm)
  return(V)

# Generate the 1PDMs and the DFT integrator 
mf=scf.UHF(m)
mf.kernel()
P = mf.make_rdm1() 
h0=mf.get_hcore()
md=dft.UKS(m,xc=xc)
ni = md._numint

# Function to evaluate HF, PBE, PBE+U, PBE+USIC, and PBE+UGVB energies of a given 1PDM 
def GetEs(P):
  Jmat2 = md.get_j(dm=P)
  Jmat = Jmat2[0] + Jmat2[1] 
  K = md.get_k(dm=P)
  Eother = numpy.einsum('sij,ij->',P,h0) + numpy.einsum('sij,ji->',P,Jmat)/2. 
  EHF = Eother - numpy.einsum('sij,sji->',P,K)/2. 
  n, exc, vxc = ni.nr_uks(mol=m,grids=md.grids,xc_code=xc,dms=P)
  EDFT = Eother + exc 
  Pproj = ProjDM(P)
  EU = Uval*( -1.0*numpy.einsum('sij,sji->',Pproj,Pproj)+numpy.einsum('sii->',Pproj))
  EDFTU = EDFT + EU 
  (pvalsa,pvecsa)=numpy.linalg.eigh(Pproj[0])
  (pvalsb,pvecsb)=numpy.linalg.eigh(Pproj[1])
  #print('pvalsa :',pvalsa)
  #print('pvalsb :',pvalsb)
  EUSIC2 = 0 
  EUGVB0 = 0 
  DeltaAB0 = 0 
  EcAB0 = 0
  EcAB1 = 0
  EXex   = 0 
  EXCdft   = 0 
  for p in range(nproj):
    na = pvalsa[p]
    nb = pvalsb[p]
    EXthis = -J*(na**2+nb**2)
    EXCdftthis =  EXCp(na,pvecsa[p],nb,pvecsb[p])
    #DeltaAB1 = numpy.abs(na+nb-1.) # Correct where projection state is exactly 1s orbital 
    DeltaAB1 = numpy.abs(na+nb-0.96596512)
    EXex  = EXex + EXthis 
    EXCdft= EXCdft + EXCdftthis
    
    EUSIC2 = EUSIC2 + EXthis -EXCp(na,pvecsa[p],0,pvecsb[p])-EXCp(0,pvecsa[p],nb,pvecsb[p])
    EUGVB0 = EUGVB0 + EXthis -EXCdftthis
    if(na>0.000000001 and nb>0.00000001):
      EcAB0 = EcAB0 + J*(DeltaAB0-(DeltaAB0**2+4*na*nb*(1-na)*(1-nb))**0.5)
      EcAB1 = EcAB1 + J*(DeltaAB1-(DeltaAB1**2+4*na*nb*(1-na)*(1-nb))**0.5)
  EDFTUGVB0 = EDFT + EUGVB0 + EcAB0 
  EDFTUGVB1 = EDFT + EUGVB0 + EcAB1 
  # Returns HF, DFT, DFT+U, DFT+USIC, DFT+UGVB0, DFT+UGVB1, , HFX, PBEXC,GVBC0, GVBC1
  return [EHF,EDFT,EDFTU,EDFT+EUSIC2,EDFTUGVB0,EDFTUGVB1, EXex,EXCdft,EcAB0,EcAB1]

v1 = GetEs(P)
Pcore = P[1]
Pd = P[0] -P[1] 
summary=''

# Generate the fractional occupancies na, nb 
fracs=[]
val=0
while(val<1.0001):
  fracs.append(val)
  val = val +0.01
#val = 0.2 
#while(val<1.0):
#  val = val +0.05
#  if(val<1.0):
#    fracs.append(val)

# Evaluate the energy at each na,nb
for na in fracs:
  #for nb in fracs:
    nb = na 
    Pf = numpy.zeros_like(P)
    Pf[0] = Pcore+na*Pd
    Pf[1] = Pcore+nb*Pd
    v2 = GetEs(Pf)
    #summary = summary+'%.2f,%.2f,%4f,%4f,%4f,%4f,%4f,%4f,%.4f,%.4f,%.4f,%.4f\n' % (na,nb,v2[0],v2[1],v2[2],v2[3],v2[4],v2[5],v2[6],v2[7],v2[8],v2[9])
    summary = summary+'%.3f,%.3f,%5f,%5f,%5f,%5f,%5f,%5f,%.5f,%.5f,%.5f,%.5f\n' % (na,nb,v2[0],v2[1],v2[2],v2[3],v2[4],v2[5],v2[6],v2[7],v2[8],v2[9])
  #summary=summary+'\n'
print(summary) 
\end{verbatim}}

%\clearpage\newpage 
%\section{Additional References} 
%\bibliographystyle{aip}
%\bibliography{../../../review/tosubmit/tosubmit}
\end{document}